\documentclass[twocolumn,prl]{revtex4}
\usepackage{epsfig,amssymb,amsmath}
\usepackage[english]{babel}
\begin{document}
\title{Shapiro Step on the rc-Branch in the IV-characteristic of a Josephson Junction with an $LC$-shunting}
\author{Yu. M. Shukrinov~$^{1,2}$}
\author{I. R. Rahmonov~$^{1,3}$}
\author{K. V. Kulikov~$^{1,2}$}
\author{P. Seidel$^{4}$}

\address{$^{1}$ BLTP, JINR, Dubna, Moscow Region, 141980, Russia \\
$^{2}$ International University of Dubna, Dubna,  141980, Russia\\
$^{3}$Umarov Physical Technical Institute, TAS, Dushanbe, 734063 Tajikistan\\
$^{4}$Institut f$\ddot{u}$r Festk$\ddot{o}$rperphysik, Friedrich Schiller Universit$\ddot{a}$t, Jena, D-07743 Jena, Germany
}

\date{\today}

\begin{abstract}
The effect of external radiation on the phase dynamics of Josephson junctions shunted by an LC-circuit is examined. It is shown that additional resonant circuit (rc) branches appear in the current-voltage characteristics of the junctions when the Josephson frequency is equal to the frequency of the formed resonance circuit. We show that the amplitude dependence of the Shapiro step width crucially changes when the Shapiro step is on the rc-branch in comparison to the case of Josephson junction without shunt. The experimental implementation of these effects might give very important advantages for existing  methods and technologies.
\keywords{Microwave radiation, nonequilibrium,  Shapiro step, intrinsic Josephson junctions.}
\end{abstract}
\maketitle

One of the effective methods for affecting the Josephson junction (JJ) is its shunting by LCR elements \cite{hadley87,wiesenfeld96,filatrella00,grib02,fistul07}. In particular, shunting leads to the synchronization of oscillations of the superconducting current in array of JJs. JJs, together with the LCR elements, form an oscillatory circuit. When the Josephson frequency $\omega_J$ becomes equal to the natural frequency of the circuit $\omega_{rc}$, oscillations in  JJ are tuned to this frequency. This resonance is manifested in the current--voltage characteristic in the
form of various features such as steps \cite{jensen90,larsen91}, humps or dips \cite{tachiki11,zhou09}. In particular, the existence of steps in the current--voltage characteristics in various systems of JJs with a resonance circuit was reported in a number of experimental and theoretical
works (see \cite{likharev86-eng,almaas02}, and references therein). A peak in the intensity of coherent electromagnetic radiation from a two--dimensional system of JJs based on Nb/Al/AlOx/Nb was detected in \cite{barbara99} at the synchronization of oscillations in different JJs, which is caused by this resonance. The considered system has an interesting potential for applications in quantum metrology~\cite{hriscu2013}.

An interesting problem concerns the effect of external electromagnetic radiation. First, as in the general case (without shunts), the radiation leads to the appearance  of the Shapiro steps (SS) and their subharmonics in the IV-characteristics \cite{hamilton00}, whose the position and width of which depend on the radiation frequency and amplitude. Second, double resonances can be realized at the conditions when  $\omega_J=(m/n)\omega_{rc}=(p/q)\omega_R$, where $\omega_J$, $\omega_{rc}$, $\omega_{R}$ are
the Josephson, resonance circuit and radiation frequency, and $m,n,p,q$ are integer numbers. The properties of SS in this case are not investigated yet. Up to now there is no study of SS on a $rc$-branch. Particularly, the question of the amplitude dependence of the width of SS on the $rc$-branch has no answer today.

In this Letter, we  study numerically the influence of the external electromagnetic radiation on the phase dynamics of JJ shunted by LC-elements. A very interesting phenomenon is found: When the corresponding SS is on the rc-branch, the amplitude dependence of its widths demonstrates the new feature in comparison with JJ without shunting. It is determined by the effective frequency which varies in a wide interval. The properties of SS on the $rc$-branch  offer promising applications.

\begin{figure}[htb]
 \centering
\includegraphics[width=42mm]{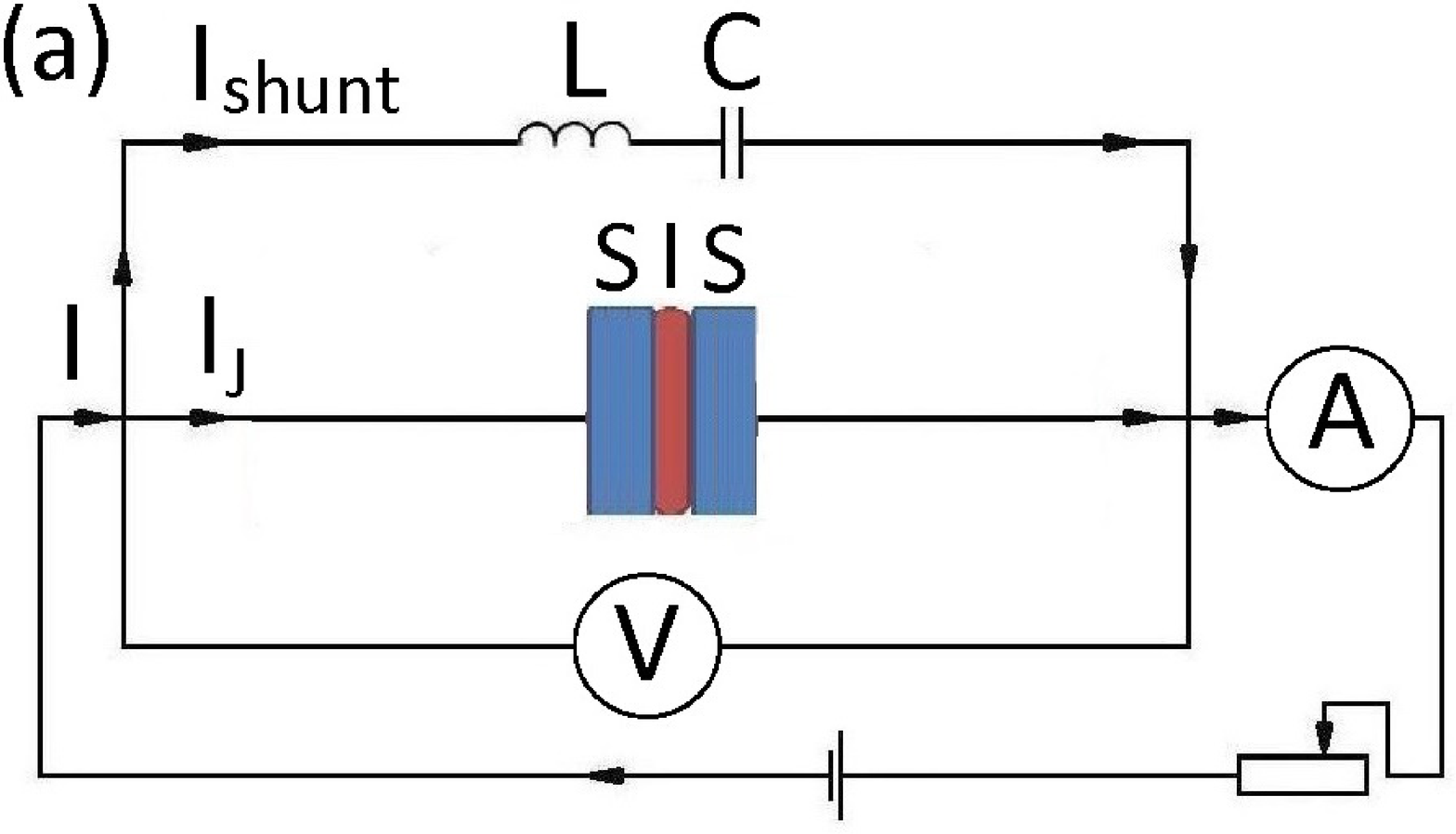}\hspace{0.3cm}\includegraphics[width=40mm]{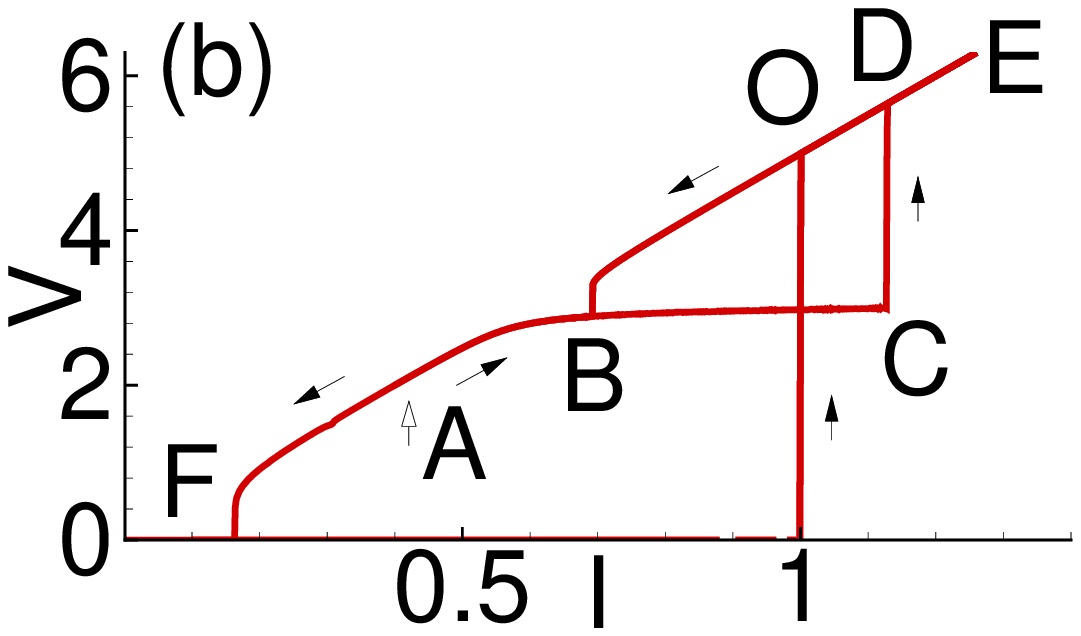}
\caption{(Color online) (a) Scheme of JJ with the $LC$ -- shunting elements; (b) The IV-characteristics of the shunted JJ, simulated by up and down sweeping of the bias current. The arrows indicate the direction of sweeping. There is an rc -- branch between points A and C.}  \label{1}
\end{figure}

The scheme of JJ with the $LC$ shunting elements is presented in  Fig.~\ref{1}(a).
In normalized units the system of equations describing this electric scheme can be written in the form of the usual RCSJ model for shunted JJ \cite{shukrinov-jetpl12-rahmonov-eng}
\begin{equation}
\label{system_eq3}
\left\{\begin{array}{ll}
\displaystyle\frac{\partial \varphi}{\partial t}=V
\vspace{0.3 cm}\\
\displaystyle \frac{\partial V}{\partial t}=I-\sin \varphi-\beta\frac{\partial \varphi}{\partial t}-C\frac{\partial u_{c}}{\partial t}+ A\sin\omega_R t + I_{noise}
\vspace{0.3 cm}\\
\displaystyle \frac{\partial^{2} u_{c}}{\partial t^{2}}=\frac{1}{LC}\bigg(V-u_{c}\bigg)
\end{array}\right.
\end{equation}
Here the bias current $I$ is normalized to the critical current $I_c$ of JJ, time $t$ - to the inverse plasma frequency $\displaystyle\omega_{p}=\sqrt{\frac{2eI_{c}}{C_{j}\hbar}}$, the voltages $V$ and $u_c$ (voltage at the capacitance $C$) normalized to $\displaystyle V_{0}=\frac{\hbar \omega_{p}}{2e}$;  shunt capacitance $C$- to the capacitance of the JJ $C_j$, and shunt inductance $L$ - to $(C_{j}\omega_{p}^{2})^{-1}$. In the system of equations (\ref{system_eq3}) we introduce a dissipation parameter $\displaystyle\beta=\frac{1}{R_{j}}\sqrt{\frac{\hbar}{2eI_{c}C_{j}}}=\frac{1}{\sqrt{\beta_{c}}}$ with $\beta_{c}$ as the McCumber parameter. Amplitude $A$ and frequency $\omega_{R}$ of external radiation  are normalized, respectively to the $I_{c}$ and $\omega_{p}$.  To reflect the experimental situation,  we have added the noise in bias current $I_{noise}$, which is produced  by random number generator and it's amplitude is normalized  to the critical current value $I_c$. Here we present the results for $\beta=0.2$. We note that JJ together with LC-elements form a resonance circuit with its eigenfrequency
\begin{eqnarray}
\omega_{rc}=\sqrt{\frac{1+C}{LC}}
\label{w}
\end{eqnarray}

First, we shortly discuss the properties of the shunted JJ in the absence of external radiation. As it was demonstrated in Refs.\onlinecite{tachiki11,zhou09}, the LC shunting leads to the step structure in the one--loop IV-characteristics. However, that step is actually a branch \cite{shukrinov-jetpl12-rahmonov-eng}, which can be totally restored by changing the direction of bias current sweeping. The corresponding IV-characteristic is presented in Fig.~\ref{1}(b). It is calculated by solving the system of equations (\ref{system_eq3}) by sweeping the bias current along $01OEBABCDEBF0$. The $IV$-characteristic has a  $rc$-branch $AC$ as a result of the resonance  of the Josephson  and eigen mode oscillations of the formed resonance circuit. The chosen shunt parameters $L=0.2$ and $C=1.25$ lead to the eigenfrequency of the resonance circuit $\omega_{rc}=3$ calculated by formula (~\ref{w}), so the exact equality $\omega_{J}=\omega_{rc}$ is fulfilled at the end of the $rc$-branch (at $I=I_{end}$, point $C$), where $V=3$. We see a jump of voltage in the IV-characteristic at this point. The form of the obtained $rc$--branch depends on the parameters of the LC circuit. It demonstrates a gentle slope in the voltage scale. The variation of the  JJ and resonance circuit parameters changes the curvature of this slope~\cite{shukrinov-jetpl12-rahmonov-eng}.

\begin{figure}[htb]
 \centering
 \includegraphics[width=78mm]{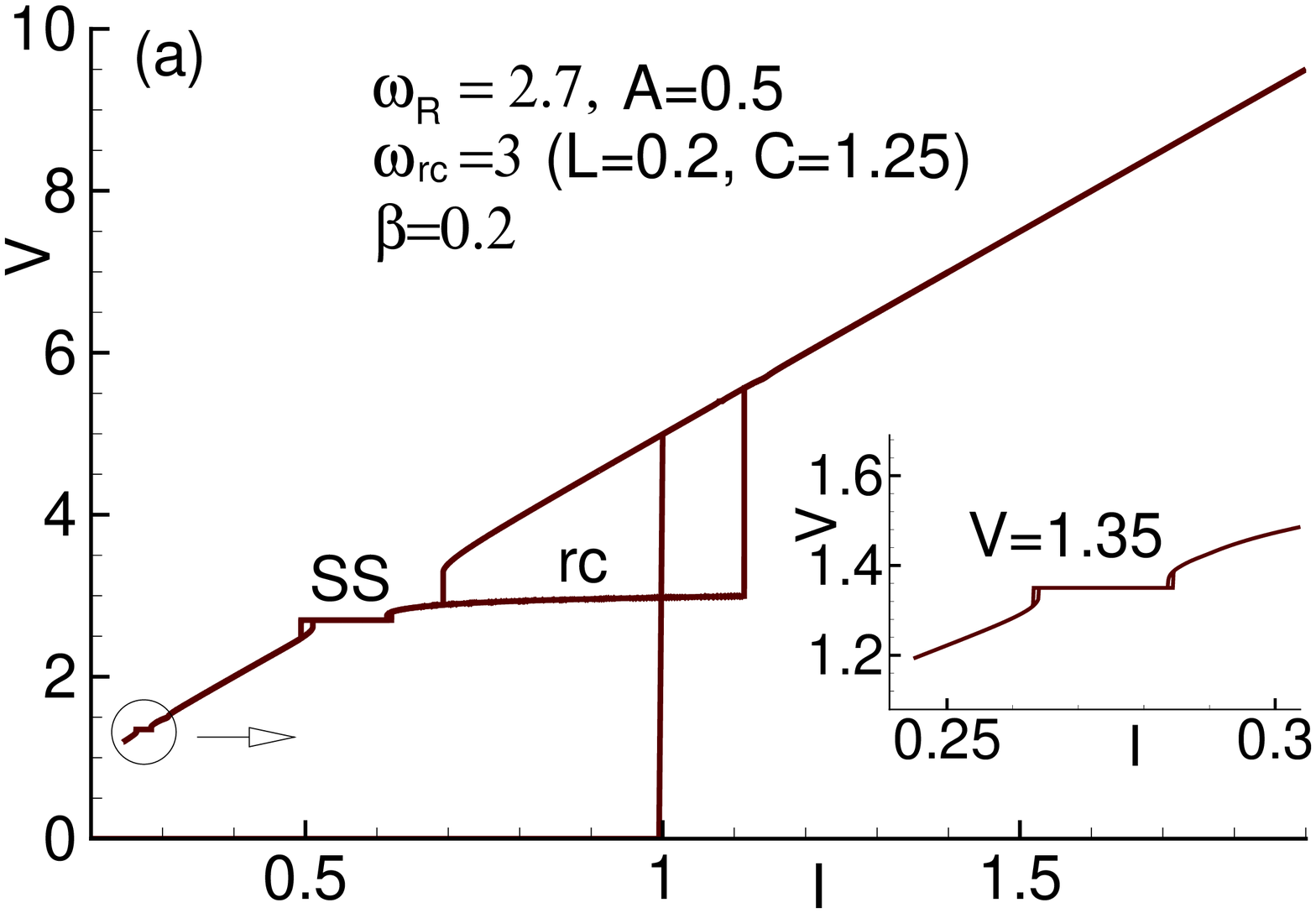}
 \includegraphics[width=78mm]{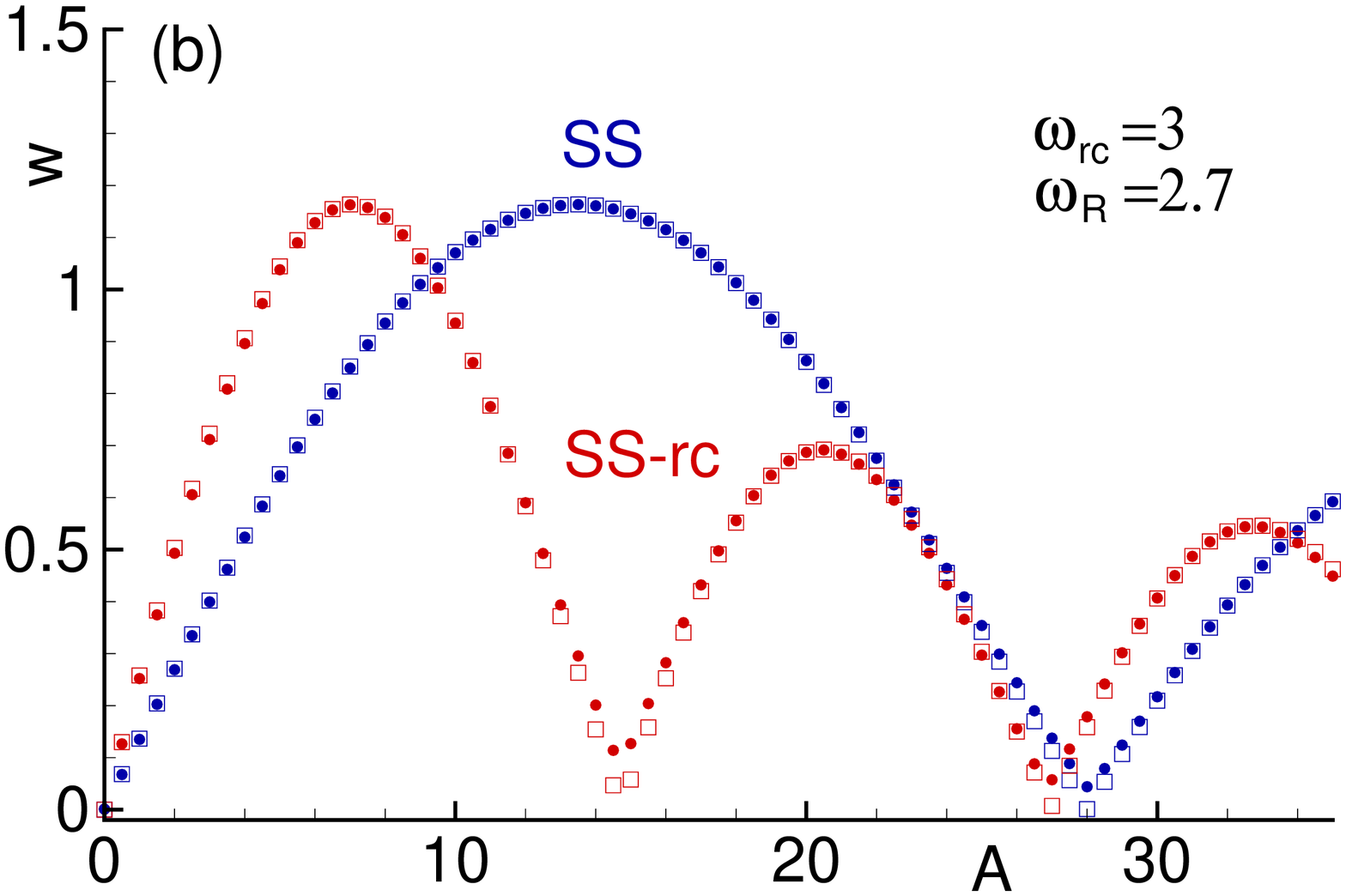}
\caption{(Color online) (a) Effect of radiation with the frequency $\omega_R=2.7$ and amplitude $A=0.5$ on the IV-characteristic of the shunted JJ. The inset enlarges the part marked by the circle demonstrating SS subharmonic; (b) The amplitude dependence of the width of SS on the rc-branch (SS--rc) at $\omega_R=2.7$, compared to the dependence without shunting (SS). Squares and dots are results of simulations and theory (\ref{step_width_e}), respectively.}  \label{2}
\end{figure}

\begin{figure}[htb]
 \centering
  \includegraphics[width=78mm]{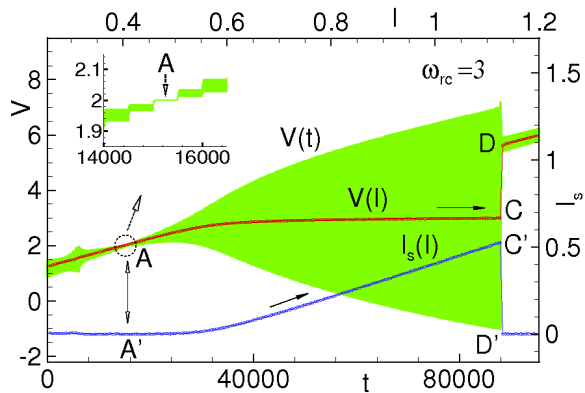}
\caption{(Color online) Time dependence of the JJ voltage together  with the corresponding part of the IV-characteristic and $I_s(I)$ calculated with an increase in the bias current. The arrows indicate the direction  of current sweeping. The characteristic points of $I_s(I)$ dependence are marked by letters with prime. The insert demonstrates a disappearance of Josephson oscillations at point $A$. A step like behavior of voltage on time is a numerical effect related to the jumps in bias current in the simulations.}  \label{4}
\end{figure}

Let us now discuss the effect of external electromagnetic radiation. This effect is produced by the term
$A\sin\omega_R t$ in the system of equations  (\ref{system_eq3}) and depends on
the values of the frequency $\omega_R$ and amplitude $A$. Figure \ref{2}(a) presents part of the IV-characteristics at   $A=0.5$ and $\omega_{R}=2.7$. As we see, in comparison with Fig.~\ref{1}(b) the IV-characteristic demonstrates additionally a Shapiro step at $V= \omega_R=2.7$ and its subharmonic at $V=\omega_R/2=1.35$. According to the results presented in  Fig.\ref{4}(a), the main SS is at the beginning of the rc-branch.

The width of SS in the absence of shunting is determined by~\cite{barone82,seidel91}
\begin{equation}
\Delta I=2|J_{n}(Z)|, \hspace{0.5cm} Z=\frac{A}{\omega_R}\frac{1}{\sqrt{\beta^2+\omega_R^{2}}}
\label{step_width_e}
\end{equation}
\noindent where $J_{n}$ is the Bessel function of the $n$--th order. The argument  $Z$ depends on the frequency and amplitude of external radiation and the parameter of dissipation  $\beta$. In Fig.~\ref{2}(b) we show the dependence of the SS width on $A$ at $\omega_R=2.7$ in both cases: with and without shunting. The results marked by SS  demonstrate the A-dependence of the SS  width in the case without shunting: The symbols are related to the  simulation results and the curves plot the theoretical dependence obtained by eq.(\ref{step_width_e}). We see that the results of simulations in the case without shunting are in good agrement with theoretical equation (\ref{step_width_e}). The results marked by the SS-rc show the dependence when SS is on the rc-branch. We see that the amplitude dependence of the Shapiro step width are crucially changed in the case of shunting. At small $A$ the width is larger in the case of shunting  than the width of SS for JJ without shunting. The period of the Bessel function is decreased in comparison with the case of the Josephson junction without shunting.

\begin{figure}[h]
 \includegraphics[width=78mm]{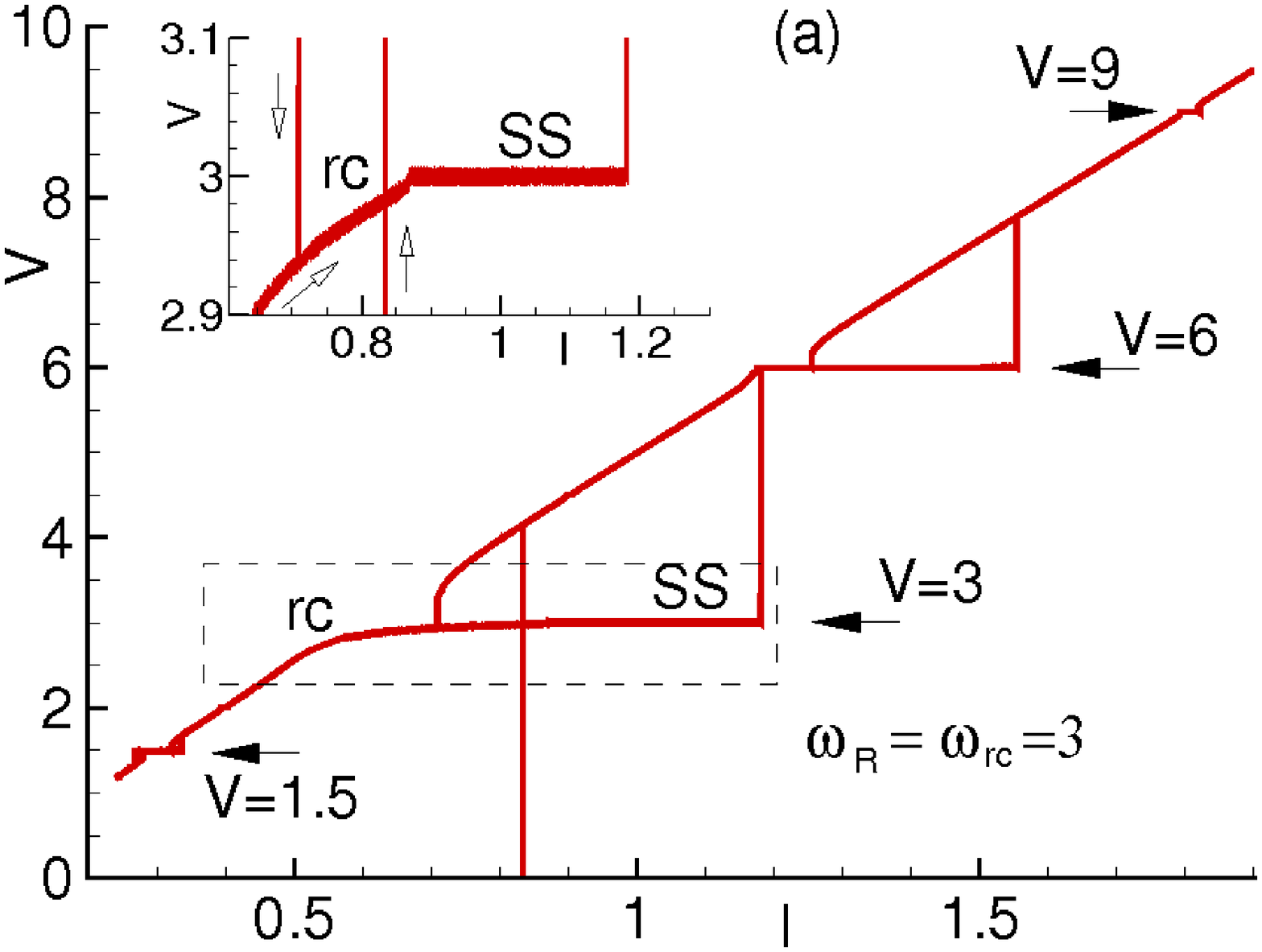}
\includegraphics[width=78mm]{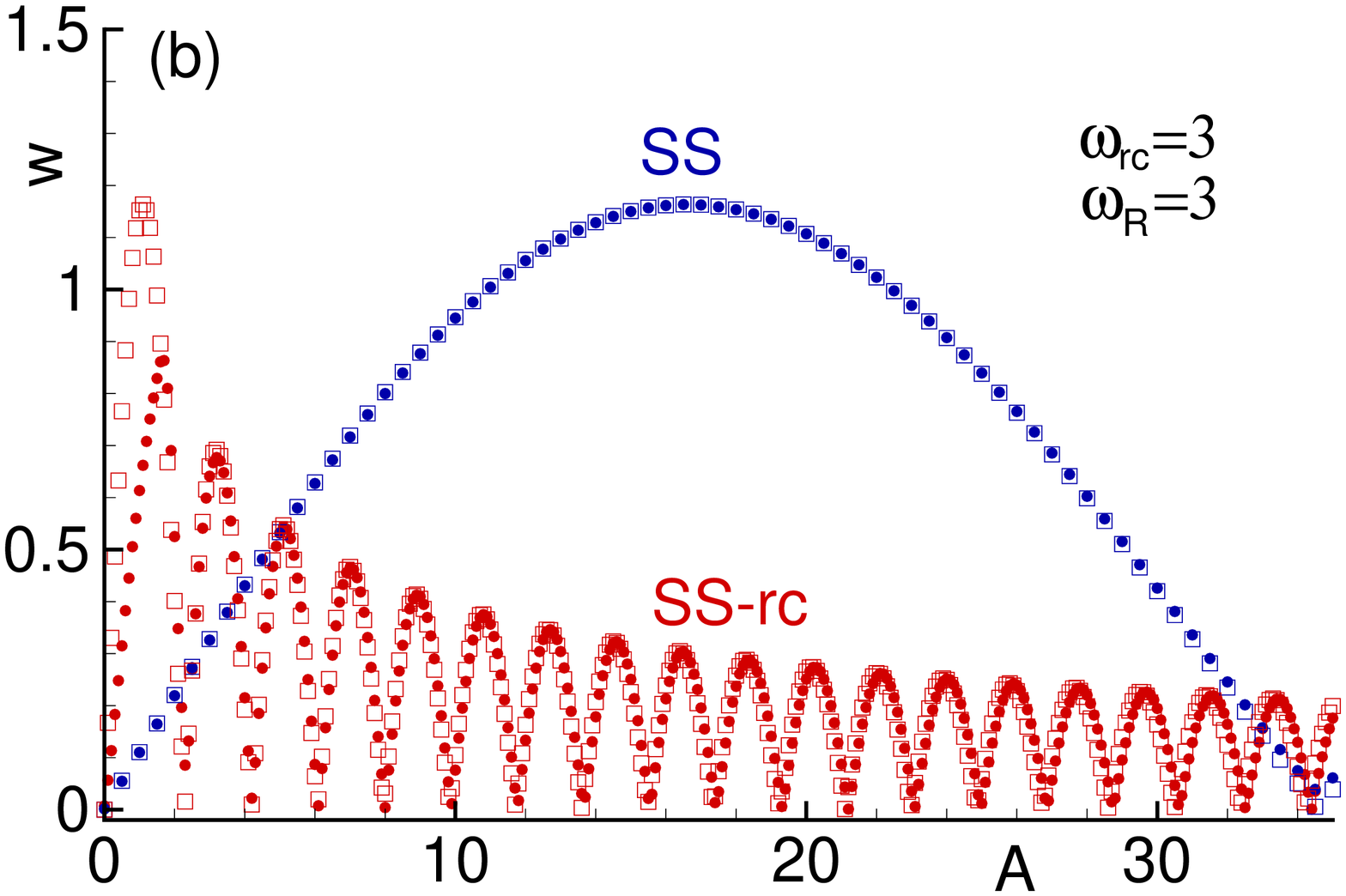}\\
\vspace{0.4cm}
\includegraphics[width=78mm]{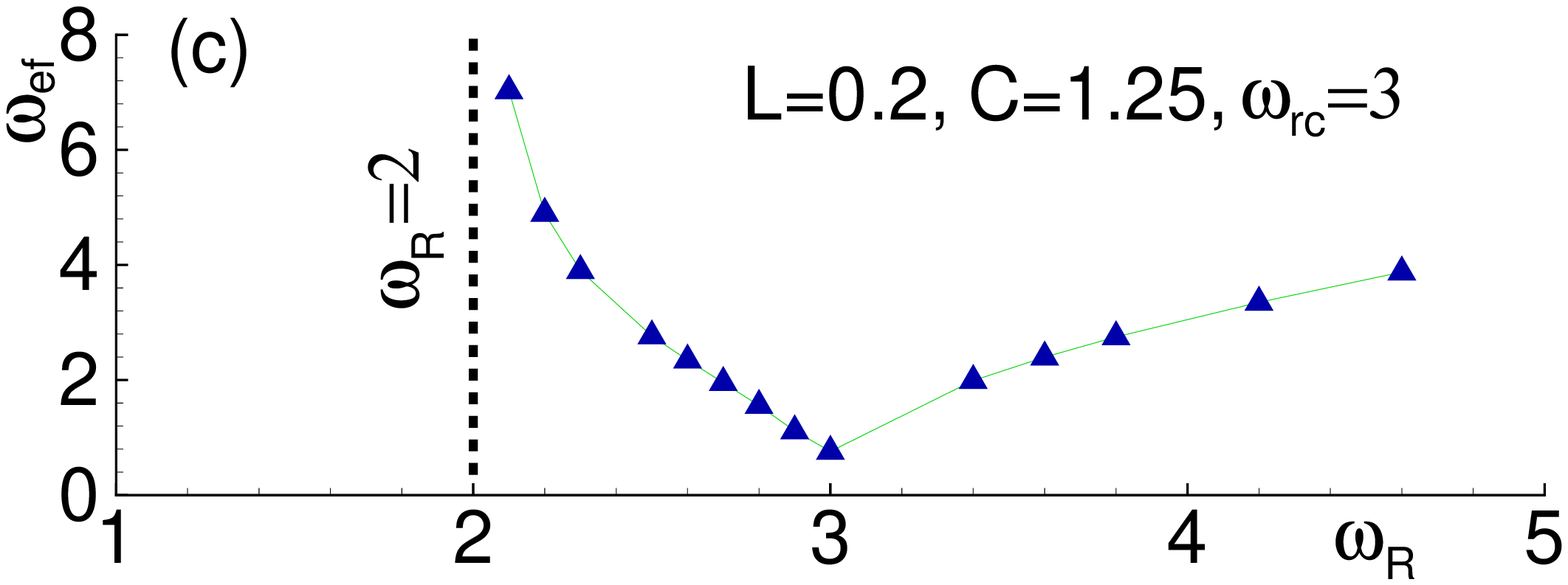}
\caption{(Color online) (a) The same as in Fig.\ref{2}(a) for $\omega_R=3$. The arrows indicate the branch at the resonance $\omega_R=\omega_{rc}=3$, harmonic at $V=6$, $V=9$ and subharmonic at $V=1.5$. The inset enlarges the part marked by dashed rectangular demonstrating SS on the rc-branch. The arrows in the inset show the direction of bias current sweeping; (b) The same as in Fig.\ref{2}(b) for $\omega_R=3$; (c) Dependence of the effective frequency on the radiation frequency at $\omega_{rc}=3$. Dashed line stresses the dependence around onset of the $rc$-branch.}
\label{3}
\end{figure}

Approaching the radiation frequency to the frequency of resonance circuit makes the observed effect stronger. Figure \ref{3}(a) presents the IV-characteristics obtained at the frequency $\omega_{R}=3$ equal to the eigenfrequency of the resonance circuit, i.e. at the double resonance  condition $\omega_{R}=\omega_J=\omega_{rc}$. The first SS in this case is on the top of the rc--branch.  In the inset to Fig.~\ref{3}(a) we enlarge this part of the IV-characteristic to demonstrate clearly the step at $V=3$. We note that the double resonance triggers the appearance of other SS harmonics and subharmonics. Three of them (at $V=1.5$, $V=6$ and $V=9$) are indicated by the arrows.

The dependence of the SS width on $A$ at this frequency $\omega_R=3$ is presented in Fig.~\ref{3}(b). The effective frequency is essentially smaller in comparison with the results in Fig.\ref{2}(b).

Using (\ref{step_width_e}) for the analysis of the obtained results on the A-dependence of  the SS  width  in the case with shunting, we fit the dependence with an effective frequency $\omega_{ef}$ instead of the normalized radiation $\omega_{R}$, and find that they  correspond to the effective frequency  $\omega_{ef}=1.955$ in the case of $\omega_{R}=2.7$ and  $\omega_{ef}=0.76$ in the case of $\omega_{R}=3$. So, shunting influences the reaction of JJ on the external radiation. Instead of the frequency $\omega_{R}$  a very good fit of the power dependence can be obtained by a reduced effective frequency $\omega_{ef}$. Because the external microwave frequency is fixed, this means that the normalization is changed from the Josephson frequency to a new higher frequency describing the locking process at the resonance branch. Now the real working point at the rc branch of the IV characteristic results in a changed frequency.    For the fixed voltage corresponding to the external frequency the IV characteristic at the rc-branch is at a higher average current and an increased slope (differential resistance), compared to the IV characteristic without shunt.

To clarify these changes, we investigate the influence of the external radiation at different frequencies in the interval $2.0 \geq \omega_{R} \leq 4.6$  for $\omega_{rc}=3$  and found the effective frequency $\omega_{ef}$ in each case. The results are demonstrated in Fig.\ref{3}(c), where  $\omega_{ef}$ is shown as a function of $\omega_{R}$. We see a decrease of the effective frequency $\omega_{ef}$  until approaching the resonance condition $\omega_{R} = \omega_{rc}$, where the effective frequency has its minimum. Approaching the point $A$ in the voltage scale by radiation frequency leads to the increase of the effective frequency.
The observed phenomenon can be manifested at different resonance circuit frequency. We get qualitatively the same result in case $\omega_{rc}=4$ \cite{prb2014}.

The obtained amplitude dependence of the SS width and the origin of effective frequency variation, can be understood by the voltage-time dependence of the shunted JJ. This dependence together with the corresponding part of the IV-characteristics calculated with an increase in  bias current sweeping  is presented in Fig.\ref{4}.
The voltage-time dependence reflects the features on the IV-characteristics and  allows one to find the boundaries of the rc-branch. It shows the maximal amplitude of voltage oscillations at the end of the $rc$-branch. On the other side, we clearly see the peculiarity of the voltage-time dependence at the point $A$ where the amplitude of Josephson oscillations is close to zero. This point corresponds to the onset of the $rc$-branch. In this figure we show also the bias current dependence of the average superconducting current $I_s=\sin{\varphi}$, simulated at the same sweeping process.  The $I_s$ has a minimum  at the point $A'$. The position of the point $A'$ is in agreement with a peculiarity in the voltage-time dependence and indicates also  the onset of the $rc$-branch. We stress this fact by double arrows in Fig.\ref{4}. Along the $rc$-branch $I_s$ is growing practically linearly. After a jump from the $rc$-branch (point $C$) and transition to the resistive state, $I_s$ decreases by two orders of magnitude and then slowly goes to zero. The inset demonstrates the disappearance of voltage oscillations around the point $A$.

So, at the end of the $rc$-branch the Josephson oscillations have a maximal amplitude and this fact is a reason  for small effective frequency which has its minimum at the resonance  $\omega_{R} = \omega_{rc}$  (point C). Opposite, at the onset of the $rc$-branch the amplitude of Josephson oscillations limites to zero, which leads to the increase of the effective frequency corresponding to a normalization frequency even higher than the Josephson frequency in case without shunting.

As summary, we studied the resonance features of JJ shunted by the LC-elements under electromagnetic irradiation.   A strong effect of the external radiation on the IV-characteristics and voltage-time dependence is demonstrated. Crucial changes are found at the double resonance condition when radiation frequency coincides with the Josephson and resonance circuit frequencies. It changes the dependence of the width of SS on the amplitude of radiation. The deviations of the appearance of the Shapiro steps and their changed power dependence are of interest not only from a basic research viewpoint.  It can be very useful for experiments, too. The optimized LC shunts lead to the increased step height at low amplitudes for steps on the rc-branch. In many cases, new types of Josephson junctions were characterized by their microwave response but often the microwave power available is small. Thus, the power dependence cannot be observed in a wide range, see e.g. \cite{doh00}.  In this case, the shunting of the junction will provide an extended range using the same microwave source, because SS demonstrates the Bessel behavior at a much smaller power of radiation in comparison to the case of unshunted JJ.  These features of SS on the $rc-$ branch might be interesting for quantum metrology \cite{jeanneret09}. Additionally, the LC shunt can be adapted in the layout to act as a wave guiding structure improving the coupling to the radiation. Another example may be the detection of radiation with a single Josephson junction by observation of a Shapiro step induced by the incoming signal \cite{barone06}.  If the range of the signal frequency is known, the LC shunt can produce a branch with enhanced sensitivity in the IV-characteristic of the detector junction. As the last example, one can think about the optimization of the programmable Josephson voltage standards where only the first Shapiro step is used  \cite{hamilton00}.

This work was supported by the Heisenberg-Landau Program. I.R.R. thanks financial support of Russian Foundation for Basic Research (RFBR) under grant 13-02-90905-mob-sng-st and JINR under grant 13-302-08.

\end{document}